\documentstyle[aps,epsfig,prl,twocolumn]{revtex}

\begin{document}

\vspace{0.0cm}
\draft

\title{Comment on ``Does the rapid appearance of life on Earth suggest
that life is common in the Universe''}
\author{V.V. Flambaum \thanks{email address:
flambaum@newt.phys.unsw.edu.au}}

\address{ School of Physics, University of New South Wales,
Sydney 2052, Australia}

\maketitle

\begin{abstract}
 In a recent  paper astro-ph/0205014  Lineweaver
and  Davis performed a statistical analysis to claim that
the rapidity of biogenesis on Earth indicates high probability
of biogenesis on terrestrial- type  planets. We argue
that the rapid appearance of life on Earth hardly tells us
anything about the probability of life to appear on  other planet.
The conclusion should be different. The rapid initial biogenesis
is consistent with a large number (N of the order of 10) of crucial
steps in evolution from simplest life forms to humans.

\end{abstract}



\section{Discrete evolution model}
 Let us consider a model of ``discrete evolution''.
It is known that the evolution from the simplest life
form to humans may be expressed in terms of
 qualitative steps, for example, first live cell,..., cell with
a nucleus,..., multicell organism, ... , mollusk, fish, amphibian,
reptile, mammal, human. Let us denote the total number
of ``crucial steps'', or ``classes'' by $N$, and probability
of a new class $(n)$ to appear from a previous class $(n-1)$ per unit
time as $\lambda_n$. The corresponding probability for a class $n$ to appear
during a small time $t_n$ is $P_n=\lambda_n t_n$. The total
probability for a human to appear during time
\begin{equation}
\label{S}t=\sum_n t_n
\end{equation}
can be presented as the product of the probabilities
\begin{equation}
\label{P} P(t)=\Pi_n \lambda_n t_n= \Pi_n \lambda_n \, \cdot \, \Pi_n t_n .
\end{equation}
 The maximum of this expression is achieved for all $t_n=t/N$.
\begin{equation}
\label{Pmax} P_{max}(t)=(\frac{t}{N})^N \, \cdot \,\Pi_n \lambda_n \, .
\end{equation}
Note that the probability $P_1=\lambda_1 t_1$ for the simplest life form
 to appear can be arbitrary small. However,  the first live organism
 should appear during a short time $t_1=t/N$ if the number of crucial
steps to produce humans $N$ is large.

     Lineweaver and  Davis  \cite{1} estimate
the biogenesis time as
\begin{equation}
\label{t1} t_1=0.1 ^{+0.5} _{-0.1} {\rm Gyr} \, .
\end{equation}
By comparing this with the time life has existed on Earth
\begin{equation}
\label{t} t \simeq 4 {\rm Gyr} \, .
\end{equation}
we conclude that the number of crucial steps in evolution
from the simplest life form to humans is $N \sim 10$.

\section{The cascade model}
To make this conclusion  more reliable we present
 another toy-model that somehow takes
into account a continuous character of the evolution.
 Let us consider a direct chain of classes (steps)
which leads from the first life to humans.

  The time dependence of the  populations $W_k(t)$ in these classes
 can be determined from the following equations:
$$
\frac{dW_1}{dt}=\Gamma_1 W_0
$$
$$
.\,\,.\,\,.\,\,.\,\,.\,\,.\,\,.\,\,.\,\,.\,\,.\,\,.
$$
\begin{equation}
\label{Weq}\frac{dW_k}{dt}=\Gamma_k W_{k-1}
\end{equation}
$$ .\,\,.\,\,.\,\,.\,\,.\,\,.\,\,.\,\,.\,\,.\,\,.\,\,. $$
 The  term $\Gamma_k W_{k-1}$ in the right-hand-side
of (\ref{Weq}) is responsible for the flux from the previous
class.
In what follows we assume that the
initial conditions are $W_0(0)=1$ (no life) and $W_k(0)=0$ for $k>0$.

Equations (\ref{Weq}) have the simple solution,
\begin{equation}
\label{sol}W_n=\frac{(\Pi_n \Gamma_n) t^n}{n!}.
\end{equation}

    The probability for  life to appear and the probability of humans
to appear (class $N >> 1 $) in this model is determined
by arbitrary parameters $\Gamma_n$.  However, for any $\Gamma_n$
 class $N$ is populated long after class 1.
Indeed, to reach the same probability $W_N(t_N)=W_1(t_1)$
the time should be
\begin{equation}
\label{tn}
t_N \simeq \frac{N (\Gamma_1 t_1)^{1/N}}{e \Gamma}
\end{equation}
Here $\Gamma=(\Pi_n \Gamma_n)^{1/N}$ is the geometrical average value.

  We can slightly modify this model by writing an equation
describing the probabilty of the system to reach a certain class
during its evolution. For simplicity we assume that all $\Gamma_n=\Gamma$.
 For this case the probabilities  for
different classes can be determined by the ``probability
conservation equations",
$$
\frac{dW_0}{dt}=-\Gamma W_0 $$
$$
\frac{dW_1}{dt}=\Gamma W_0-\Gamma W_1
$$
$$
.\,\,.\,\,.\,\,.\,\,.\,\,.\,\,.\,\,.\,\,.\,\,.\,\,.
$$
\begin{equation}
\label{Weq1}\frac{dW_k}{dt}=\Gamma W_{k-1}-\Gamma W_k
\end{equation}
$$ .\,\,.\,\,.\,\,.\,\,.\,\,.\,\,.\,\,.\,\,.\,\,.\,\,. $$
 The first term $\Gamma W_{k-1}$ in the right-hand-side
of (\ref{Weq1}) is responsible for the flux from the previous
class, and the second term $\Gamma W_k$ describes the transition
from the class $k$, into the next class $k+1$.

Equations (\ref{Weq1}) have the simple solution, $$ W_0=\exp
(-\Gamma t\,) $$
\begin{equation}
\label{sol1}W_n=\frac{(\Gamma t)^n}{n!}\exp \left( -\Gamma t\right) =\frac{
(\Gamma t)^n}{n!} W_0.
\end{equation}

 For an infinite chain one can easily
check the normalization condition for the probabilities,
\begin{equation}
\label{norm}\sum\limits_{n=0}^\infty W_n=\exp (-\Gamma
t)\,\sum\limits_{n=0}^\infty \frac{(\Gamma t)^n}{n!}=1.
\end{equation}

The maximal probability $W_n=\frac{n^n}{n!}\exp (-n)\approx
1/\sqrt{2\pi n}$ to be in the class $n$ determined by the
condition $\frac{dW_n}{dt}=0$ , occurs for $t=n/\Gamma $ ,
therefore, this solution (\ref{sol}) can be considered as a
{\it cascade} in the population of different classes .
Indeed, at small times $t\ll \tau \equiv 1/\Gamma $ the system is
practically in the initial state (no life), at times $t\approx \tau \,$ the
flow spreads into the first class, for $t=n\tau $ it spreads into
the $n-$th class, etc.

    The probability for  life to appear and the probability of humans
to appear (class $N >> 1 $) in this model is determined
by a single parameter $\Gamma$.  However, again
 class $N$ is populated long after class 1.  For  $ t << 1/\Gamma$
this model gives the same result as a  model considered above.

\section{Conclusions}
 The probabilities of the first life to appear
are determined by the arbitrary parameters $\lambda_1$ in the first
model and $\Gamma$ in the second model.
For any values of these paramers the first life should appear on a
 much shorter time scale than ``humans'' (if the number of steps $N>>1$).

  Thus, the rapid appearance of life on Earth hardly tells us
anything about the probability of life to appear on another planet.
It only tells us that there was a large number of crucial
intermediate steps between the first live organisms and humans.

\section{Acknowledgments}

This work was supported by the Australian Research Council.
I am  grateful to Paul Curmi and Michael Kuchiev  for valuable comments.

\end{document}